\documentclass[aps,prb,twocolumn,amsmath,amssymb,floatfix,superscriptaddress]{revtex4-2}
\usepackage{blindtext}
\usepackage{epsfig}
\usepackage{color}
\usepackage{amsmath}
\usepackage{amsfonts}
\usepackage{amssymb}
\usepackage{graphicx}%
\usepackage{url}
\usepackage[breaklinks]{hyperref}
\usepackage[english]{babel}
\usepackage{esvect}
\usepackage[normalem]{ulem}
\usepackage[inline]{enumitem}
\usepackage{xcolor}
\usepackage{xfrac}
\usepackage{dsfont}
\usepackage{bm}
\usepackage{array}
\newcolumntype{C}[1]{>{\centering\let\newline\\\arraybackslash\hspace{0pt}}m{#1}}
\usepackage{footnote}
\usepackage{isomath}
\DeclareMathAlphabet\mathbfcal{OMS}{cmsy}{b}{n}

\hypersetup{colorlinks,citecolor=blue,filecolor=blue,linkcolor=blue,urlcolor=blue}

\bibpunct{[}{]}{,}{n}{}{}

\begin{document}

\title{Bipartite Fluctuations and Charge Fractionalization in Quantum Wires}

\author{Magali Korolev and Karyn Le Hur}
\affiliation{CPHT, CNRS, École polytechnique, Institut Polytechnique de Paris, 91120 Palaiseau, France}
\date{\today}

\begin{abstract}
We introduce a quantum information method for measuring fractional charges in ballistic quantum wires generalizing bipartite fluctuations to the chiral quasiparticles in Luttinger liquids, i.e. 
analyzing and summing charge and current fluctuations in a region of the wire. Bipartite fluctuations at equilibrium are characterized through a logarithmic scaling with distance encoding the entangled nature of these fractional charges in one-dimensional (1D) fluids. This approach clarifies the physical meaning of the dephasing factor of electronic interferences in a ballistic ring geometry at zero temperature, as a result of charge fractionalization. 
We formulate an analogy towards ground-state energetics. We show how bipartite current fluctuations represent a useful tool to locate quantum phase transitions associated to Mott physics. We address a spin chain equivalence and verify the fractional charges through an algorithm such as Density Matrix Renormalization Group (DMRG). Adding a potential difference between the two sides (parties) of the wire, bipartite fluctuations can detect a bound state 
localized at the interface through the Jackiw-Rebbi model coexisting with fractional charges. 
\end{abstract}

\maketitle

\section{Introduction}

In 1D quantum systems, electronic interactions give rise to unique and paradigmatic phenomena \cite{Haldane,Giamarchi}. The ground-state wavefunction of a Luttinger liquid has a Jastrow form similar to the Laughlin wavefunction \cite{PhamOrsay,Laughlin}. Low-dimensional ballistic quantum wires then are characterized by charge fractionalization similar to the fractional quantum Hall physics, but chiral quasiparticles can move in both directions, left or right \cite{PhamOrsay,SafiSchulz}. 
It is yet delicate to measure those fractional charges. The conductance of a quantum wire reaches $\frac{e^2}{h}$ as for free electrons, 
assuming spin-polarized electrons, which can be understood as a Fabry-Perot resonance of fractional wavepackets \cite{SafiSchulz,MaslovStone,Tarucha}. Signatures of charge fractionalization can be found through momentum-resolved tunneling when injecting an electron with fixed chirality in the middle of a wire \cite{Steinberg,KarynBertAmir}; this measure reveals the difference or Asymmetry between the (fractional) charges moving in different directions which is proportional to the Luttinger parameter. 
Interferometry measures in a ballistic wire allows us to access information on dephasing effects from interactions, e.g. at finite temperature the dephasing time reveals  the associated electron lifetime in a Luttinger liquid \cite{Karyn,Rice}. A dephasing time proportional to the inverse of temperature was reported in ballistic rings of GaAs \cite{Hansen}, but both electron-electron interactions \cite{Karyn} and dephasing from a surrounding metallic gate \cite{SeeligButtiker} can explain the signal. In chiral edge states of fractional quantum Hall systems various tools are developed and proposed to measure the fractional charges  with great accuracy \cite{Kapfer} such as current noise at zero frequency \cite{Glattli,Picciotto} and Ramsey interferometry in time \cite{TalKaryn}. 

To motivate the present work, we emphasize that direct measures of the entangled nature of many-body systems are possible in a laboratory
through quantum information theory tools such as bipartite entanglement entropy \cite{Lin,Harvard}. For Luttinger liquids, this entropy shows a logarithmic behavior with distance \cite{CalabreseCardy}.  The 1D Fermi gas and the Luttinger liquid reveal the same entanglement entropy. In this work, our goal is to clarify how we can reveal fractional charges of 1D quantum wires from such equilibrium methods in a quantum information perspective.
It is useful to consider charge fluctuations occurring between two macroscopic subsystems. One of us introduced this tool, through the 
 {\it bipartite charge fluctuations} \cite{chargefluctuations}, as a marker of the many-body entangled nature of ground states. Bipartite charge fluctuations precisely mean that we study the variance (noise) associated to the charge or number of electrons in one of the two regions of the wire. They show a logarithmic dependence on the system size or of time in analogy to the entanglement entropy.  This probe reveals the Luttinger parameter $K$ as a prefactor which has the significance of the compressibility from an analogy to statistical physics \cite{chargefluctuations}. To reveal the fractional charges $f_R=\frac{1+K}{2}$ and $f_L=\frac{1-K}{2}$ in units of the electron charge \cite{PhamOrsay,SafiSchulz,Steinberg,KarynBertAmir,Karyn}, in this work, we introduce the {\it bipartite chiral charge fluctuations} i.e. bipartite fluctuations of charge associated to the true chiral quasiparticles which move to the left and to the right. This will be obtained when adding (summing) {\it bipartite charge and current fluctuations}: Eqs. (\ref{chiralcharges}), (\ref{fluctuations}) and (\ref{chargefluctuations}) will reveal the fractional charges  i.e. $f_L^2+f_R^2$ as a prefactor of the logarithmic scaling. We formulate a relation to ground-state energetics in Eq. (\ref{energetics}).
 Also, we clarify the link with the dephasing of mesoscopic interferences at zero temperature in Eq. (\ref{formuladephasing}). 

From Landauer's point of view, noise here is a quantum signal i.e. it precisely measures the entangled nature of chiral quasiparticles. Bipartite charge fluctuations of chiral quasiparticles in the fractional quantum Hall effect are also a useful tool to probe edge states \cite{Fradkin,FQHE1/2}. It is also interesting to mention that noise measures are attracting attention in the community,  theoretically and experimentally, related to low-dimensional quantum systems \cite{Berg,Gutman,Biswas}.

\section{Luttinger liquid and Fractional Charges}
\label{Luttingerliquid}

To reach these goals, in this Section, we find it useful to introduce the model and to remind a few facts associated to fractional charges in ballistic quantum wires.
The Luttinger Hamiltonian corresponds to the low-energy model associated to interacting fermions hopping on a lattice through adjacent sites \cite{Haldane,Giamarchi}.

The Luttinger liquid Hamiltonian reads
\begin{equation}
H = \frac{u}{2\pi}\int_0^L dx \left( \frac{1}{K}(\partial_x \phi)^2 + K (\partial_x \theta)^2 \right).
\label{Luttinger}
\end{equation}
Here, $u$ is the sound velocity and $K$ is the Luttinger parameter such that $K=1$ for free electrons and for interacting electrons $K<1$. The charge density reads
$\rho=\rho_0 -\frac{1}{\pi}\partial_x \phi$, with $\rho_0$ the mean density of electrons, and the current density takes the form $j=\frac{uK}{\pi}\partial_x \theta$. We remind the correlation relations $[\phi(x),\nabla\theta(y)]=i\pi \delta(x-y)$.
The correspondence with the free electron gas can be reached when modifying 
$\phi\rightarrow \sqrt{K}\tilde{\phi}$ and $\theta\rightarrow \tilde{\theta}/\sqrt{K}$. 
The ground state correlation functions of the modified fields take the symmetric forms
\begin{eqnarray}
\langle (\tilde{\phi}(x) - \tilde{\phi}(0))^2\rangle = {\cal F}_1(x) \\ \nonumber
\langle (\tilde{\theta}(x)-\tilde{\theta}(0))^2\rangle = {\cal F}_1(x) 
\end{eqnarray}
with the logarithmic function 
\begin{equation}
{\cal F}_1(x) = \frac{1}{2} \log \left(\frac{\alpha^2 +x^2}{\alpha^2}\right),
\end{equation}
and the ground state correlation functions of the original fields then take the form \cite{Giamarchi}
\begin{eqnarray}
\langle (\phi(x) - \phi(0))^2\rangle = K {\cal F}_1(x) \\ \nonumber
\langle (\theta(x)-\theta(0))^2\rangle = \frac{1}{K} {\cal F}_1(x).
\end{eqnarray}
Here,  $\alpha$ represents the lattice spacing or short-distance cutoff. 
The electron operator for a {\it right} or {\it left} mover reads \cite{Giamarchi}
\begin{equation}
c_{R/L}(x) = \frac{1}{\sqrt{2\pi \alpha}} e^{i(\mp \phi + \theta)}.
\label{electronoperator}
\end{equation}
Right and Left movers refer to the direction of propagation along the wire.
From results above, then we reveal the general form for the electron Green's function 
\begin{equation}
\langle c_R(x) c^{\dagger}_R(0)\rangle = \frac{1}{2\pi \alpha} e^{-\frac{1}{K}\left(f_L^2 + f_R^2\right){\cal F}_1(x)}.
\label{Greenfunction}
\end{equation}
To understand the physical meaning of the fractional charges $f_L$ and $f_R$ in this equation, we remind here that the Luttinger liquid Hamiltonian can be equivalently written as \cite{PhamOrsay,Karyn}
\begin{equation}
H = \frac{uK}{4\pi}\int_0^L dx\left((\partial_x \theta_R)^2 + (\partial_x \theta_L)^2\right),
\end{equation}
where $\theta_{R,L} = \theta\mp \frac{\phi}{K}$. The fractionalization mechanism can then be understood simply saying that these fields associated to left and right modes in the wire are different than the
fields $\mp \phi+\theta$ defining the electron operators in Eq. (\ref{electronoperator}). Therefore, a right or left-moving electron will decompose itself into fractional quasiparticles. To reveal these fractional quasiparticles in Eq. (\ref{quasiparticles})
we proceed as follows. We introduce the associated charge variables $\phi_{R,L}=\frac{1}{2}(-\phi\pm K\theta)$ such that $\phi=-(\phi_R+\phi_L)$ and $\theta=(\phi_R-\phi_L)/K$.
They satisfy the commutation relations $[\theta_{R/L}(x),\partial_y \phi_{R/L}(y)]=i\pi\delta(x-y)$. Integrating the chiral charge densities $\frac{1}{\pi}\partial_x \phi_{R,L}$ then we precisely reveal the selection rules for the fractional charges:
\begin{equation}
\label{chargesgeneral}
\frac{1}{\pi}\int_0^L \partial_x \phi_{R,L} dx = \frac{Q\pm K \tilde{J}}{2}=Q_{R,L}.
\end{equation}
We have introduced the (integrated) charge $Q=-\frac{1}{\pi}\int_0^{L}\partial_x\phi(x) dx$ and the current $\tilde{J}=\frac{1}{\pi}\int_0^L \partial_x\theta dx$.
It is then useful to introduce the integrated chiral charges such that $Q_L+Q_R=Q$ and $J=u(Q_R-Q_L)=uK\tilde{J}$. We have the identifications for the {\it charge} $Q$ and {\it current} $\tilde{J}$:
\begin{eqnarray}
Q(L) &=& -\frac{1}{\pi}(\phi(L)-\phi(0)) \nonumber \\
\tilde{J}(L)&=& \frac{1}{\pi}(\theta(L)-\theta(0)).
\end{eqnarray}
If we add an electron e.g. from a reservoir such that $Q=+1$ and $J=uK$ with the identity $uK=v_F$ for one wire \cite{Giamarchi} i.e. $\tilde{J}=+1$ then we verify the occurrence of fractional charges $Q_R=\frac{1+K}{2}=f_R$ and $Q_L=\frac{1-K}{2}=f_L$ \cite{SafiSchulz,PhamOrsay,KarynBertAmir}. 
In this way, the electron operator can be equivalently introduced as \cite{Karyn}
\begin{equation}
\label{quasiparticles}
\psi_R(x) = \frac{1}{\sqrt{2\pi \alpha}}{\cal L}_R^{f_R} \cdot {\cal L}_L^{f_L}
\end{equation}
where 
\begin{equation}
{\cal L}_{R,L}^{f_R,f_L} = e^{i f_{R,L} \theta_{R,L}}.
\end{equation}

We emphasize that from momentum-resolved tunneling, it is possible to measure the difference between the fractional charges, $f_R-f_L=K$, from DC current or transport measures \cite{Steinberg,KarynBertAmir}. The electron Green's function in real space is not accessible in an experiment, therefore in Sec. \ref{chargesfluctuations} we provide several complementary views of Eq. (\ref{Greenfunction}) associated to charge and current fluctuations which may lead to further practical applications, e.g. as an interpretation of dephasing at zero temperature in a quantum wire through the physical meaning of {\it entangled fractional charges}. Related to this aim, in Sec. \ref{chargesfluctuations} we propose below to measure the factor $\frac{1}{K}\left(f_L^2 + f_R^2\right){\cal F}_1(x)$ through bipartite {\it charge} and {\it current} fluctuations within the ground state at equilibrium. For free electrons, the function ${\cal F}_1$ precisely enters in the formula for bipartite charge fluctuations which acquires a nice interpretation in terms of entangled measure between two regions of the quantum fluid, a region $A$ of length $x$ and the region $B$ of length $L-x$ \cite{chargefluctuations}. Therefore, for interacting electrons, as we develop below, the factor $\frac{1}{K}\left(f_L^2 + f_R^2\right){\cal F}_1(x)$ will acquire a clear understanding in terms of entangled fractional charges measurable through the dephasing profile of mesoscopic interferences in 1D ballistic rings at zero temperature i.e. in the quantum limit. 

In Sec. \ref{chargeDMRG} we show through DMRG that bipartite current fluctuations are precisely measurable numerically e.g. in quantum spin chain analogues. In this way, it allows us to reveal the presence of fractional charges in 1D quantum fluids.
It is also a very precise tool to locate quantum phase transitions e.g. associated to Mott physics. 
In Sec. \ref{JackiwRebbi}, we elaborate on the idea that bipartite current fluctuations can probe the presence of a localized bound state at a (topological) interface through the Jackiw-Rebbi model coexisting with fractional charges.

\section{Fractional Charges, Bipartite Fluctuations and Dephasing in a Ring}
\label{chargesfluctuations}

From the identifications between integrated charge and current in a domain of size $x$, which can be understood as
$Q(x)=-\frac{1}{\pi}(\phi(x)-\phi(0))$ and $\tilde{J}(x)=\frac{1}{\pi}(\theta(x)-\theta(0))$, we can evaluate bipartite charge and current fluctuations associated to this region. We emphasize here that this region is referred to as region $A$ and the complementary region of length $L-x$ will be e.g. the region $B$.

Bipartite charge fluctuations in the region of the wire with $x\gg \alpha$ (in the quantum limit $x\ll \beta u$ where $\beta=\frac{1}{k_B T}$) read
\cite{chargefluctuations}
\begin{eqnarray}
\label{one}
{\cal F}_{Q(x)} &=& \langle Q(x)^2\rangle - \langle Q(x)\rangle^2 = \frac{K}{\pi^2}\ln \frac{x}{\alpha} \\ \nonumber
&=&\sum_{i,j\in [0;\frac{x}{\alpha}] }(\langle n_i n_j \rangle - \langle n_i\rangle \langle n_j\rangle),
\end{eqnarray}
with 
\begin{equation}
Q_A = Q(x)=\sum_{i=0}^{\frac{x}{\alpha}} c^{\dagger}_i c_i =  \sum_{i=0}^{\frac{x}{\alpha}} n_i.
\end{equation}
We introduce  the electron creation (annihilation) operator on the lattice $c^{\dagger}_i$ $(c_i)$ and the corresponding density $n_i = c^{\dagger}_i c_i$. From the definitions above, we also have $n = n_L+n_R$ at position $i$ along the wire with $n_L$ and $n_R$ referring to the density of electrons moving to the left and to the right. In this formula, the measure is done at any time $t$ within the ground state. In the continuum limit, we equivalently have the relations for the charge and current densities $\rho(x)=n_L(x)+n_R(x)$ and $j(x)=u(n_R(x)-n_L(x))$. The symbol ${\cal F}_Q(x)$ can be thought of in a symmetric way i.e. ${\cal F}_{Q(x)}={\cal F}_{Q_A(x)}={\cal F}_{Q_B(L-x)}$ when the total charge $Q_A+Q_B$ is fixed. Bipartite charge fluctuations is related to the quantum Fisher information e.g. in the Kitaev p-wave superconductor \cite{Herviou} and Hubbard model \cite{Konik}. The quantum Fisher information is measured in quantum spin chains through the dynamical spin susceptibility \cite{Tennant}. 

Here, we introduce bipartite fluctuations associated to the current $\tilde{J}(x)$ (which were not introduced before) from the fact that the correlator in $\theta$ also shows a logarithmic scaling:
\begin{eqnarray}
\label{fluctuationscurrent}
{\cal F}_{\tilde{J}(x)} &=& \langle \tilde{J}(x)^2\rangle - \langle \tilde{J}(x)\rangle^2 = \frac{1}{K\pi^2}\ln \frac{x}{\alpha} \nonumber \\
&=& \sum_{i,j\in A}(\langle \tilde{j}_i\tilde{j}_j \rangle - \langle \tilde{j}_i\rangle \langle  \tilde{j}_j\rangle),
\end{eqnarray}
with the corresponding current density on the lattice $j_i = it(c^{\dagger}_i c_{i-1} - c^{\dagger}_{i-1} c_i)$ where $\tilde{j}_i=j_i/t$ for the 1D fermions, $\tilde{J}(x)=\tilde{J}_A=\sum_{i\in A} \tilde{j}_i$ and $t$ the hopping strength (related to the hopping term); the Planck constant $\hbar=\frac{h}{2\pi}$ is set to unity. Hopping of electrons on the lattice results in an Hamiltonian $H=-t\sum_i (c^{\dagger}_i c_{i+1}+h.c.)$ and the current density is obtained from the Heisenberg equation of motion $\frac{d}{dt} n_i = i [H,n_i]$. The interaction term commutes with the density operator. In the continuum limit, $\tilde{j}(x)=\frac{1}{K}(n_R(x)-n_L(x))$. Equivalently, ${\cal F}_{J(x)} =  \langle J(x)^2\rangle - \langle J(x)\rangle^2 = \frac{v_F^2}{K\pi^2}\ln \frac{x}{\alpha}$. The measure again is done at any time (usually, $t$ designates both the hopping amplitude and the time). Within our definitions, we can verify the identity $\frac{\partial}{\partial t}Q(x,t) = -\frac{\partial}{\partial x}J(x,t)$.

To reveal the chiral quasiparticles, we can then introduce the total charge in the region $A$ associated to left-moving and right-moving particles, $Q_L=\sum_{x'\in A} n_L(x')$ and $Q_R=\sum_{x'\in A} n_R(x')$, which are respectively related to $\phi_L$ and $\phi_R$ variables. Then, Eqs. (\ref{one}) and (\ref{fluctuationscurrent})
lead to
\begin{equation}
\label{chiralcharges}
{\cal F}_{Q_L(x)} = {\cal F}_{Q_R(x)} = \langle Q_L^2(x)\rangle - \langle Q_L(x)\rangle^2 = \frac{K}{2\pi^2}\ln\frac{x}{\alpha}.
\end{equation}
This is also interpreted as 
\begin{equation}
\frac{{\cal F}_{Q(x)}+{\cal F}_{K{\tilde{J}}(x)}}{4}={\cal F}_{\frac{Q(x)\pm K\tilde{J}}{2}}={\cal F}_{Q_R}(x)={\cal F}_{Q_L}(x).
\end{equation}
From Eq. (\ref{chargesgeneral}), the chiral charge quasiparticles operators are $Q_{R/L}=\frac{(Q\pm K\tilde{J})}{2}$ such that the equation above refers to the bipartite fluctuations associated
to those fractional charges. Fluctuations associated to the chiral quasiparticles in Eq. (\ref{chargesgeneral}) then reveals the Luttinger parameter $K$. 
The identity  $\frac{{\cal F}_{Q(x)}+{\cal F}_{K{\tilde{J}}(x)}}{4}={\cal F}_{\frac{Q(x)\pm K\tilde{J}}{2}}$ assumes that $\langle \phi(x)\theta(0)\rangle \rightarrow constant$ when $x\gg \alpha$ \cite{Giamarchi}.
To precisely reveal the presence of fractional charges through $f_L$ and $f_R$, we propose to add charge and current bipartite fluctuations associated to the system of interacting electrons.
From Eqs. (\ref{one}) and (\ref{fluctuationscurrent}), then we obtain
\begin{equation}
\pi^2\frac{({\cal F}_{Q(x)}+{\cal F}_{\tilde{J}(x)})}{2} = \frac{1}{K}(f_R^2+f_L^2){\cal F}_1(x).
\label{fluctuations}
\end{equation}
\begin{equation}
\label{chargefluctuations}
\pi^2 K^2\frac{({\cal F}_{Q(x)}+{\cal F}_{{\tilde{J}}(x)})}{2} = (f_R^2+f_L^2) ({\cal F}_{Q_L}(x)+{\cal F}_{Q_R}(x)).
\end{equation}
The existence of fractional charges in Sec. \ref{Luttingerliquid} was derived from equivalences through charge and current conservations in a wire \cite{SafiSchulz,PhamOrsay,KarynBertAmir}. Therefore, to reveal the presence of those charges in the fluctuations, we show here that this also requires to sum charge and current bipartite fluctuations in a domain of the wire. When we sum these fluctuations of charge and current for the interacting electrons system this is also equivalent to measure bipartite charge fluctuations associated to the two species (colors, L or R) of chiral quasiparticles which carry a fractional charge and reveal $\frac{1}{K}(f_L^2+f_R^2)=\frac{1}{2}(K+K^{-1})$ as a prefactor. The logarithmic function ${\cal F}_1$ then can be seen as an entanglement measure of fractional charges.  This is then another way to interpret the profile of the Green's function of the electron operator in Eq. (\ref{Greenfunction}) in space, that will be important to understand the presence of dephasing at zero temperature in a ballistic ring hereafter. We also deduce
\begin{eqnarray}
\label{energetics}
\frac{\pi^3 u}{4} \frac{\partial}{\partial x} ({\cal F}_Q(x)+{\cal F}_{\tilde{J}}(x)) &=& \frac{\pi u}{2K}\frac{1}{x}(f_L^2+f_R^2)\mid_{x=\frac{L}{2}} \nonumber \\
&=& \frac{\pi u}{K L}(f_R^2+f_L^2).
\end{eqnarray}
This is precisely the energy to add the pair of fractional charges in the middle of the quantum wire (with $x=\frac{L}{2}$). 

Here, we show the potential of Eq. (\ref{fluctuations}) to measure bipartite fluctuations associated to mesoscopic interferences in a ring geometry. The signal detected from mesoscopic interferences can be written as the transmission probability \cite{Karyn} ${\cal T}=1-(T_0+T_L)+\sqrt{T_0 T_L}(e^{i\pi \varphi+iK_1-iK_2}+h.c.)$; $T_0$ and $T_L$ represent the transmission probabilities to enter the mesoscopic ring at $x=0$ and to escape from the ring at $x=L_r$, $\varphi=\Phi/\Phi_0$ is the magnetic flux with
$\Phi_0$ the flux quantum. The phases $K_1$ and $K_2$ describe the dephasing from an electrical circuit with a resistance \cite{SeeligButtiker} and also the dephasing from interactions in the Luttinger liquid \cite{Karyn}. When we perform an ensemble of measures, this phase factor measures voltage fluctuations along one path $j$ of the ring through the identity $\langle e^{i{\cal K}_j} \rangle =e^{-\frac{\langle{\cal K}_j^2\rangle}{2}}$ \cite{Karyn}. The dephasing factor corresponds to voltage fluctuations integrated along a region of the ring \cite{Karyn}
\begin{equation}
\langle {\cal K}_j^2 \rangle =\frac{e^2}{\hbar^2}\int_0^{L_r} dx' \int_0^{L_r} dx'' \langle \delta V_j(x') \delta V_j(x'')\rangle. 
\end{equation}
For Luttinger liquids, we can then rephrase the result at zero temperature in Ref. \cite{Karyn} as a correspondence with bipartite fluctuations in Eq. (\ref{chargefluctuations})
\begin{equation}
\label{formuladephasing}
\frac{1}{2}\langle {\cal K}_j^2(L_r) \rangle = \pi^2\frac{({\cal F}_{Q(L_r)}+{\cal F}_{{\tilde{J}}(L_r)})}{2} - {\cal Y}
\end{equation}
where ${\cal Y}$ is the same quantity in the absence of interactions for $K=1$ i.e. $\pi^2\frac{({\cal F}_{Q(L_r)}+{\cal F}_{{\tilde{J}}(L_r)})}{2} - {\cal Y}=\left(\frac{1}{K}(f_R^2+f_L^2)-1\right)\ln \frac{L_r}{\alpha}=2\gamma\ln \frac{L_r}{\alpha}$. Here, $2\gamma$ has the physical interpretation of a dimensionless resistance $r=\frac{R e^2}{h}$ and $x=L_r$ is the size of each path forming the ring. The dimensionless resistance comes from the analogy between the effect of intrinsic interactions and the
noise coming from a metallic gate with a resistance $R$ \cite{SeeligButtiker,Karyn}. For ballistic quantum wires, $\frac{h}{e^2}$ is the unit of resistance justifying the expression `dimensionless resistance' above. The form of fractional charges and the Luttinger parameter are measurable through the fit of this formula. 
The power-law form of mesoscopic interferences i.e. of $\langle {\cal T}\rangle$ is then a marker of the entangled nature of the two counter-propagating fractional wavepackets at zero temperature. When the right-moving charge (wavepacket) $f_R=\frac{1+K}{2}$ is measured at $x=L_r+\epsilon$ (with $\epsilon\rightarrow 0$) in the region $B$ then it is entangled with the left-moving fractional charge (wavepacket) which has moved backward and therefore is yet in the region $A$ inside the ring. The power-law decay of the interferences can be understood as the probability for the left-fractional charge $f_L=\frac{1-K}{2}$ to reach $x=L_r\sim ut$ at time $t$ i.e. $|\langle {\cal L}_{L}^{f_L}(x,t){\cal L}_L^{f_L}(0,0)\rangle|_{x=L_r\rightarrow ut}^2\propto \left(\frac{L_r}{a}\right)^{-2\gamma}$ in the electron Green's function \cite{Karyn}.

Eq. (\ref{formuladephasing}) equivalently provides a direct measure of bipartite fluctuations in quantum wires. We emphasize here that the measure of the dephasing factor as a function of length is accessible in a laboratory \cite{Hansen}.

\section{Fractional Charges and Quantum Spin Chains}
\label{chargeDMRG}

From the Jordan-Wigner transformation at half-filling, i.e. for $\langle n_i\rangle=\frac{1}{2}$, we present a correspondence for the electron current density operator in quantum spin chains leading then to the measures of fractional charges through DMRG algorithms. For spin polarized fermions with an interaction between nearest neighboring sites, the spin Hamiltonian is an XXZ chain
\begin{equation}
H =-2t\sum_{j=1}^N (S_j^x S_{j+1}^x + S_j^y S_{j+1}^y)+U\sum_{j=1}^N S_j^z S_{j+1}^z,
\end{equation}
with the length of the wire $L=N\alpha$. We have the correspondences $S_i^z = c^{\dagger}_i c_i -\frac{1}{2}$, $S_j^+ S_{j+1}^- = c^{\dagger}_j c_{j+1}$. The hopping term of fermions on the lattice becomes an $XY$ spin interaction (with equal couplings along $X$ and $Y$) and the interaction $U$ gives rise to an Ising interaction. When $U<2t$, the low-energy theory of this quantum spin model at half-filling is therefore the Luttinger liquid in Eq. (\ref{Luttinger}) \cite{Giamarchi,Affleck}. The electron charge density $n_i$ corresponds to the spin variable $S_i^z$, $n_i=S_i^z+\frac{1}{2}$ with $S_i^z=\pm \frac{1}{2}$ i.e. $\langle n_i\rangle=\frac{1}{2}$. The electron current density operator then precisely corresponds to 
\begin{equation}
j_i = t\tilde{j}_i = 2t(S_i^x S_{i-1}^y - S_i^y S_{i-1}^x)=2t{\cal O}_{i,i-1}^{xy-yx}.
\end{equation}
We develop on this correspondence in Appendix \ref{AppendixA}.
The ${\cal O}$ operator coupling sites $i,i-1$ is the analogue of the $\tilde{j}_i = i(c^{\dagger}_i c_{i-1} - c^{\dagger}_{i-1} c_i)$ current density introduced above. In the Letter \cite{chargecurrent}, the authors introduce a representation of the current which can be written in terms of Majorana operators. We develop the correspondence with the  ${\cal O}_{i,i-1}^{xy-yx}$ operator in Appendix \ref{AppendixA}.

For free fermions i.e. $U=0$, generalizing the derivations in Refs. \cite{chargefluctuations}, we find the exact formula from the $XY$ spin chain for bipartite current fluctuations 
\begin{eqnarray}
{\cal F}_{\tilde{\cal J}(x)} &=&\sum_{i,j\in A} \langle {\cal O}_{i,i-1}^{xy-yx} {\cal O}_{j,j-1}^{xy-yx}\rangle - \langle {\cal O}_{i,i-1}^{xy-yx} \rangle \langle {\cal O}_{j,j-1}^{xy-yx}\rangle \nonumber \\
&=& \frac{1}{\pi^2}\left(\ln \frac{x}{\alpha} +\gamma +\ln 2 + \frac{1}{2}\right).
\end{eqnarray}
For the sake of clarity, we present the derivation of this formula in Appendix \ref{AppendixA}, where we also remind the formula obtained for the charge fluctuations from the correlations of $S_i^z$ i.e. ${\cal F}_{Q(x)} = \frac{1}{\pi^2}(\ln \frac{x}{\alpha} +\gamma +\ln 2+1)$ \cite{chargefluctuations}. Adding interactions, this is identical to rescale the quantum fields as $\phi(x)\rightarrow \sqrt{K}\tilde{\phi}(x)$ and $\theta(x)\rightarrow \frac{1}{\sqrt{K}}\tilde{\theta}(x)$ in Eq. (\ref{Luttinger}) compared to the free electron model such that we reproduce the leading form of the correlation functions in Eq. (\ref{fluctuationscurrent}). From a correspondence with Bethe Ansatz, we have the identification for the Luttinger parameter $K = \frac{\pi}{2\arccos(-\frac{U}{2t})}$ \cite{Nishimoto}. 

\begin{figure}[t]
\includegraphics[width=9cm]{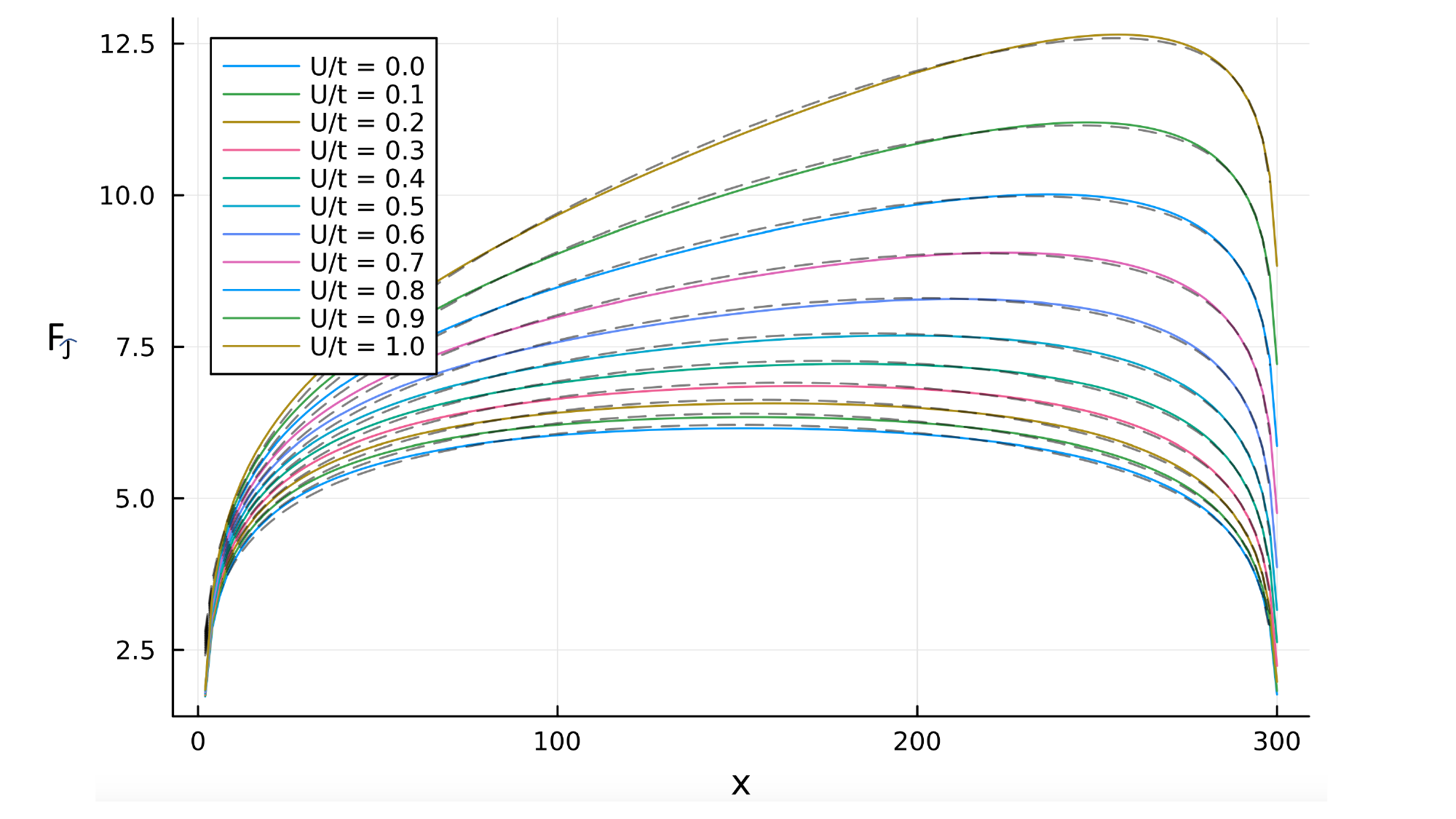}
\caption{Current fluctuations from DMRG with PBC for a wire with 300 sites, 20 sweeps. Data are in solid colored lines and the dashed lines represent a fit with a logarithmic term plus a linear term modulo a constant (see Appendix \ref{AppendixA}).
At weak $U$, results agree with the analytical form in Eq. (\ref{fluctuationscurrent}) and when $U$ becomes substantial a linear form develops revealing
the Mott transition.}
\label{Fcurrent}
\end{figure}

\begin{figure}[]
\includegraphics[width=9cm]{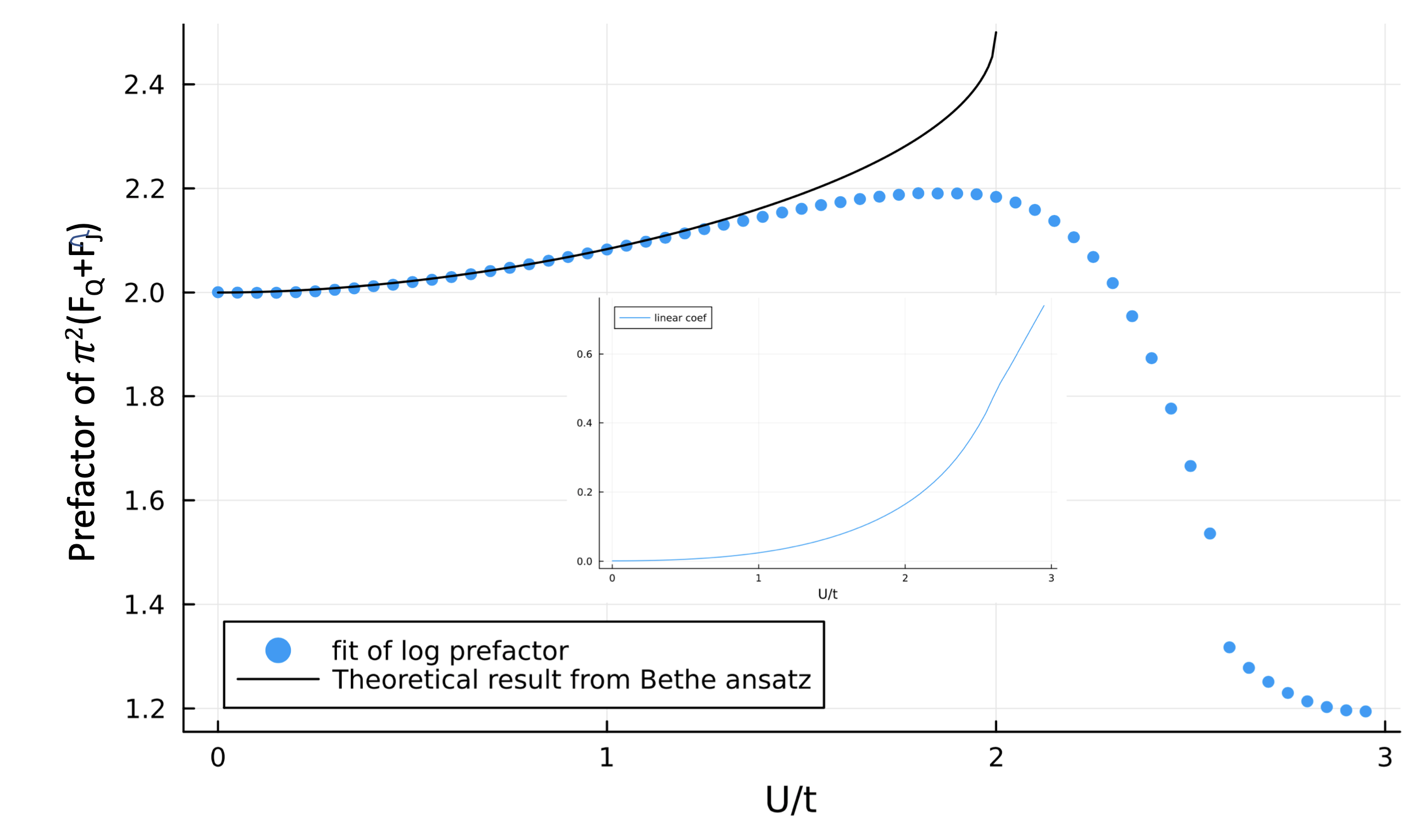}
\caption{Prefactor of the logarithmic term in $\pi^2({\cal F}_Q(x)+{\cal F}_{\tilde{J}}(x))$ with DMRG, same protocol as Fig. 1 and fit with the formula revealing the fractional charges in Eq. (\ref{fluctuations}) with precisely $\frac{2}{K}(f_R^2+f_L^2)=(\frac{1}{K}+K)$; the function $K$ in terms of $U/t$ is obtained from Bethe Ansatz \cite{Nishimoto}. (Inset) Increase of linear term at the Mott transition $(U=2t)$.}
\label{Fig2chargesfit}
\end{figure}

In Figs. \ref{Fcurrent} and \ref{Fig2chargesfit}, we present DMRG results of the formula (\ref{chargefluctuations}) with the quantum spin chain
Hamiltonian at half-filling \cite{ITensors,datajustification}. For Periodic Boundary Conditions (PBC) charge fluctuations for a subsystem of size $x$ reveal a logarithmic behavior for $U/t=0$ up to and beyond $\sim 2$. We find the Luttinger parameter in Eq. (\ref{one}) \cite{chargefluctuations}. For Open Boundary Conditions (OBC) the charge fluctuations reveal an additional Friedel oscillating term \cite{chargefluctuations}. We also verify the Luttinger parameter from the logarithmic term in the current fluctuations which is obtained from a precise protocol (see Appendix \ref{AppendixA}); current fluctuations show an additional linear term which reveals the Mott transition in Fig. \ref{Fig2chargesfit} inset. This method leads to the fit of fractional charges when adding the logarithmic terms of ${\cal F}_{\tilde{\cal J}}$ and ${\cal F}_{\cal Q}$ in Fig. \ref{Fig2chargesfit}. We also show a comparison with Bethe Ansatz.


In Ref. \cite{fluctuationsT}, we have shown that bipartite charge fluctuations are useful to study quantum phase transitions e.g. associated to Mott physics. 
We emphasize here that the bipartite current fluctuations are also a powerful and precise tool to detect the transition into a Mott phase associated to the decay of the logarithmic term of ${\cal F}_{\tilde{\cal J}}$ in Fig. \ref{Fig2chargesfit}. 
This occurs when $U>2t$ or $K<\frac{1}{2}$ such that the interaction term $S_j^z S_{j+1}^z$ in the spin language which is 
proportional to $(c^{\dagger}_R c_L)^2+h.c.$ in the quantum field theory, with the electron operators introduced in Eq. (\ref{electronoperator}), gives rise
to the relevant Sine-Gordon term $\cos(4\phi\sqrt{K})$ \cite{Affleck}, i.e. the mode $\phi(x)$ becomes fixed to a uniform value associated to this insulator i.e. a classical antiferromagnet in the spin language. The Heisenberg spin chain corresponds to $U=2t$ i.e. $K=\frac{1}{2}$. Fluctuations in the charge sector remain very small for $U>2t$ (see Appendix \ref{AppendixA}). 
The electron Green's function decreases exponentially in space $\sim \exp-\frac{x}{\xi}$ as a result of the charge density wave formation and the linear term in the bipartite current fluctuations $\frac{x}{\xi}$ i.e. give an access to the correlation length $\xi$ or inversely the energy gap. A fit with inset of Fig. \ref{Fig2chargesfit} gives the estimate $\frac{\alpha}{\xi}\sim e^{\frac{U}{t}1.4533}$. 

\section{Localization effect from the interface and Jackiw-Rebbi topological model}
\label{JackiwRebbi}

The Jackiw-Rebbi model \cite{JackiwRebbi} is a general model of interfaces with two media showing a potential difference such that it admits a zero-energy bound state
in the solution of the quantum field theory \cite{Shenbook}. The zero-energy bound state occurs even if the potential is smoothly varying around the interface located at $x=\frac{L}{2}$ i.e. with a potential of the form  
$\sim \Delta\tanh\left(\frac{x-\frac{L}{2}}{\alpha}\right)$ \cite{JackiwRebbi,Angelakis}. Here, we ask two questions: can such a bound state be observable with our tool e.g. bipartite current fluctuations? Can the zero-energy solution coexist with the fractional charges of the Luttinger liquid? We show below that our analysis can lead to positive answers. Our tool can then be useful to locate interfaces as well. 
 
\begin{figure}[t]
\includegraphics[width=\linewidth]{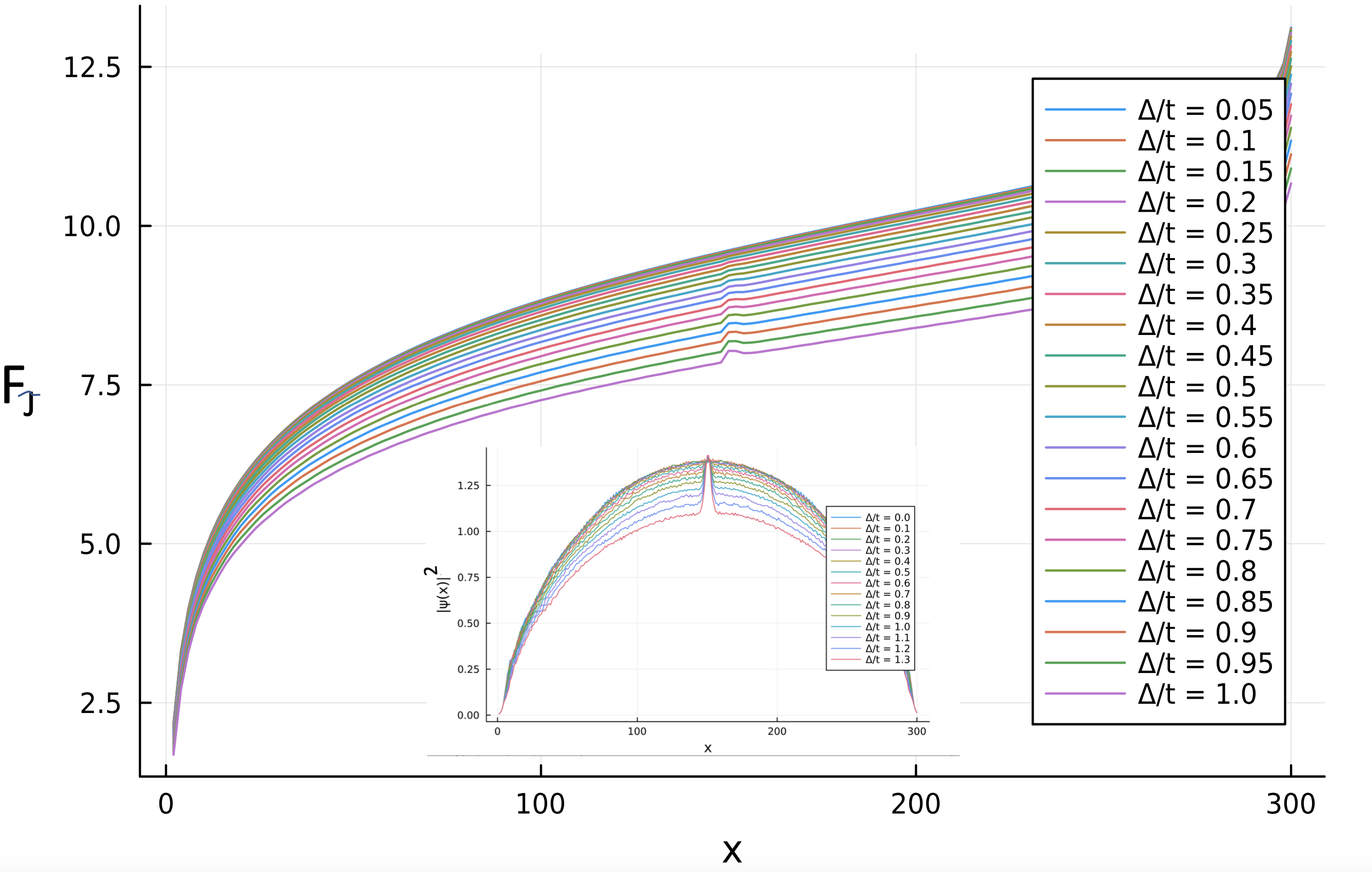}
\caption{Bipartite Current Fluctuations for OBC at $U=0.2t$ with $300$ sites, for various values of $\Delta$. (Inset) Density probability (not normalized) at small $U$ e.g. at $U=0$.}
\label{chargecurrent}
\end{figure}

We introduce the variable $y=x-\frac{L}{2}$ such that the bound state will be localized around $y=0$. 
 Looking for eigenstates resolved in energy $\epsilon$, it is useful to introduce   
\begin{equation}
c_{\epsilon}^{\dagger} = \int_{-\frac{L}{2}}^{\frac{L}{2}} dy \chi_{\epsilon}(y) c^{\dagger}(y),
\end{equation}
associated to the Hamiltonian $\sum_{\epsilon} \epsilon c^{\dagger}_{\epsilon} c_{\epsilon}$. It allows us to show how the eigenstate equation $[H,c^{\dagger}_{\epsilon}] = \epsilon c^{\dagger}_{\epsilon}$ turns into 
\begin{equation}
\label{characteristic}
\epsilon \chi_{\epsilon}(y) = -2 i v_F \partial_y \chi_{\epsilon}(y) - i2\Delta \tanh\left(\frac{y}{\alpha}\right)\chi_{\epsilon}(-y).
\end{equation}
In Appendix \ref{AppendixB}, we derive a proof of such an equation formulating a correspondence towards a quantum field theory of a 1D superconducting wire with s-wave (or d-wave) symmetry or a Luther-Emery model 
\cite{KarynEPL,LutherEmery,FabrizioGogolin}. Eq. (\ref{characteristic}) admits the zero-energy state (see Appendix \ref{AppendixB})
\begin{equation}
\chi_0(y) = \left(\frac{1}{\sqrt{\pi}}\sqrt\frac{\Delta}{2t}\right)^{\frac{1}{2}}e^{-\frac{\Delta}{2t}\ln \cosh\frac{|y|}{\alpha}}.
\end{equation}
We assume half-filling such that $v_F=2t$. First, we verify the occurrence of the bound state in the density probability and in the bipartite current fluctuations at weak $U$; see Fig. \ref{chargecurrent}. In Fig. \ref{chargecurrent} with OBC, we clearly observe a logarithmic term at small $x$ for current fluctuations, traducing the Luttinger liquid, coexisting with the bound state response at $x=\frac{L}{2}$. The prefactor of the logarithmic term decays as $\sim (1-\left(\frac{\Delta}{2t}\right)^2)$ at small to intermediate $U$, which corresponds to the probability for an electron to cross the interface or to the fractional charges to spread on the whole sample. This result shows that fractional charges can yet develop a Luttinger liquid in the presence of the interface.
For OBC charge fluctuations show pronounced oscillations as a function of distance. 

From bosonization, the interface produces a backscattering term (proportional to $c^{\dagger}_L c_R +h.c.$) i.e. $\sim\frac{\Delta}{2\pi\alpha}\tanh\left(\frac{x-\frac{L}{2}}{\alpha}\right)\cos(2\phi)$ in the Hamiltonian density (close to the interface) \cite{Giamarchi}. It goes to zero precisely when $x=\frac{L}{2}$ and for $x\neq \frac{L}{2}$ it takes a dual form as the p-wave pairing term \cite{pwavearticle} in the Kitaev p-wave superconducting wire \cite{Kitaev}, if we flip the role of the charge field $\phi$ and current field $\theta$. From this sense, this bound state can be seen as topological \cite{Shenbook}, i.e.
is stable in the presence of interactions. We emphasize here that the peaks in the density probability and in the current fluctuations remain visible for strong interactions $U\sim t$ and similarly for the logarithmic scaling at small $x$ (see 
Appendix \ref{AppendixB}). 

\section{Conclusion}

We have derived the relation between fractional charges in 1D quantum wires and bipartite chiral charge fluctuations, asumming bipartite charge and current fluctuations. This also provides a way to measure them through the mesoscopic interferences at zero temperature. We have shown a spin chain correspondence and verified the occurrence of fractional charges through DMRG. We have also shown how bipartite current fluctuations represent a useful marker to localize Mott transitions through a linear term. Introducing a smoothly varying potential at the interface between the two sides of the wire, it is yet possible to detect a localized bound state in the presence of interactions coexisting with the fractional charges. Bipartite charge, spin fluctuations and also bipartite current fluctuations may be measured from correlation functions with a double summation on space. We observe progress in developing tools e.g. in ultra-cold atoms \cite{correlations}. Similarly, coupling a metallic gate along a region of the wire would also allow to resolve information on charge dynamics similarly as in a quantum dot \cite{chargefluctuationsgate}. Our bipartite fluctuations \cite{chargefluctuations} may also be measured from the dynamical spin correlations similarly as for the measure of the quantum Fisher information \cite{Konik,Tennant}. Tools are also developed to measure fractional charges in fractional quantum Hall states \cite{Kapfer,Glattli,Picciotto,TalKaryn,Fradkin,FQHE1/2}. 
\\

Magali Korolev is grateful to Ecole Polytechnique for the kind support of her PhD thesis. The numerical results are obtained with the Cluster Cholesky at Ecole Polytechnique. 

\begin{appendix}

\section{Charge Current and Bipartite Fluctuations with the Spin Formalism}
\label{AppendixA}

Here, we derive the analogue of the electron current density in quantum spin chains and the associated bipartite fluctuations.

We have the correspondence of variables between fermions and spins $S_i^z=\pm \frac{1}{2}$ and $n_i=c^{\dagger}_i c_i=0,1$ such that $n_i=S_i^z + \frac{1}{2}$. In the bosonization language, this implies $-\nabla\phi = \pi (S^z(x) +\frac{1}{2})$ and $\sum_{i\in A} S^z_ i \rightarrow -\frac{1}{\pi}(\phi(x) - \phi(0))$ modulo a constant term. To derive the analogue of the current and current density on the lattice, we begin with the continuity equation $\nabla j_i + \partial_t n_i=0$ with $\partial_t n_i = i[H,n_i]$ taking for simplicity $\hbar=1$. This is then equivalent to $j_i = it(c^{\dagger}_i c_{i-1} - c^{\dagger}_{i-1} c_i)$, where $t$ is the hopping amplitude between adjacent sites. It is useful to remind the Jordan-Wigner correspondence between fermions and spins in one dimension
\begin{eqnarray}
S_i^+ &=& c^{\dagger}_i e^{i\pi\sum_{j<i} n_j} \\ \nonumber
S_i^- &=& e^{-i\pi\sum_{j<i} n_j}c_i.
\end{eqnarray}
From the correspondences $c^{\dagger}_{i-1}c_i=S^{+}_{i-1} S_i^-$ and $c^{\dagger}_{i}c_{i-1}=S^{+}_{i} S^-_{i-1}$, the current density takes the form
$j_i = it(S_i^{+} S_{i-1}^- - S_i^{-} S_{i-1}^+)$. This result is obtained developing the string operator $e^{\pm i \pi n_{i-1}} = 1-2n_{i-1}$ with $n_i=0,1$ and e.g. $S_i^{+} S_{i-1}^- = c^{\dagger}_i (1-2n_{i-1})c_{i-1} = c^{\dagger}_i c_{i-1}$.
From the identities between spin observables $S_i^{\pm} = S_i^x \pm i S_i^y$, we obtain the result 
\begin{equation}
j_i = 2t(S_i^x S_{i-1}^y - S_i^y S_{i-1}^x).
\end{equation}
The other terms are zero because $[S_i^x,S_{i-1}^x]=0$ and $[S_{i-1}^y,S_{i}^y]=0$. Introducing the operator ${\cal O}$ related to the electron current density $j_i = 2t{\cal O}_{i,i-1}^{xy-yx}$, we derive the precise form
\begin{equation}
{\cal O}_{i,i-1}^{xy-yx} = S_i^x S_{i-1}^y - S_i^y S_{i-1}^x = \frac{i}{2}(c^{\dagger}_i c_{i-1} - c^{\dagger}_{i-1} c_i).
\end{equation}
The subscript symbol refers to the two sites associated to the two operators defining the current density and the upperscript symbol refers to the order of the two spin operators. In the bosonization language, this is equivalent to
\begin{equation}
\sum_{i\in A} {\cal O}_{i,i-1}^{xy-yx} = \frac{1}{\pi}(\theta(x)-\theta(0)).
\end{equation}
It is interesting to observe that the two terms in the ${\cal O}$ operator have an interpretation in terms of Majorana fermions which commute with the Hamiltonian. A similar observation was done recently in Ref. \cite{chargecurrent}, where they introduce an additional phase to the hopping term. Introducing the two Majorana fermions $\gamma_{Ai}=\frac{1}{\sqrt{2}}(c_i+c^{\dagger}_i)$ and $\gamma_{Bi}=\frac{i}{\sqrt{2}}(c_i-c^{\dagger}_i)$, we indeed obtain that 
\begin{equation}
j_i = it(\gamma_{Ai}\gamma_{Ai-1} + \gamma_{Bi}\gamma_{Bi-1}).
\end{equation}

We show below the derivation of the current fluctuations from the spin chain analogue i.e. from the specific form of this operator ${\cal O}$. 
Since we will generalize the calculation done in Ref. \cite{chargefluctuations} for the charge fluctuations or spin fluctuations associated to $S_z(A)=\sum_{i\in A} S_i^z$, first we find
it useful to remind a few lines of that calculation to see the parallel between the evaluations of bipartite charge and current fluctuations from the spin chain analogy on the lattice.

To evaluate spin fluctuations associated to $S_z(A)=\sum_{i\in A} S_i^z$ corresponding to charge fluctuations for fermions in region $A$ this requires to introduce the variance
\begin{equation}
{\cal F}_{S_z(A)} = \left\langle \left(\sum_{i\in A} S_i^z\right)^2\right\rangle -  \left\langle \left(\sum_{i\in A} S_i^z\right)\right\rangle^2.
\end{equation}
For a system where the total spin magnetization is conserved in the system made of regions $A$ and $B$ then ${\cal F}_{S_z(A)} = {\cal F}_{S_z(B)}$, and
\begin{equation}
{\cal F}_{S_z(A)} = \frac{x}{4} + \sum_{i\neq j; i,j=1}^x  \langle S_i^z S_j^z \rangle.
\end{equation}
The key step is to write the spin operators in terms of fermions and apply Wick theorem such that [2]
\begin{eqnarray}
 \langle S_i^z S_j^z \rangle &=& \delta_{i,j} \langle c^{\dagger}_i c_j \rangle - |\langle c^{\dagger}_i c_j\rangle|^2 \\ \nonumber
 &=& \frac{1}{2}\delta_{i,j} - |M_{i,j}|^2.
\end{eqnarray}
From the spin correspondence $\langle c^{\dagger}_i c_i\rangle=\frac{1}{2}$ because $\langle S_i^z\rangle=0$ corresponding to half-filling. For free fermions, i.e. $U=0$, this leads to the exact result
\begin{equation}
|M_{i,j}|^2 = \frac{\sigma_{i-j}}{\pi^2(i-j)^2}.
\end{equation}
$\sigma_k=0$ if $k$ is even and $1$ if $k$ is odd. Therefore, this leads to 
\begin{equation}
\sum_{i\neq j; i,j=1}^{x} \frac{\sigma_{i-j}}{\pi^2(i-j)^2} = \frac{2}{\pi^2} \sum_{k=1}^x \frac{(x-k) \sigma_k}{k^2}.
\end{equation}
For simplicity the lattice spacing $\alpha$ is set to unity in this derivation. 
This sum is equivalent to
\begin{eqnarray}
 && \frac{2}{\pi^2} \sum_{k=1}^x \frac{(x-k) \sigma_k}{k^2} \nonumber \\
 \hskip -0.5cm &=& \frac{2}{\pi^2}\left(\frac{\pi^2}{8}x-\frac{1}{2}(\ln x+1+\gamma+\ln 2 +{\cal O}(x^{-2}))\right).
\end{eqnarray}
In this way, we verify \cite{chargefluctuations}
\begin{equation}
{\cal F}_{S_z(A)} = \frac{1}{\pi^2}(\ln x+ f_1),
\end{equation}
with $f_1=1+\gamma+\ln 2$. The logarithmic term agrees with the Luttinger liquid theory if we set $K=1$ corresponding to free fermions. 

We can then derive  the bipartite current fluctuations with the operator ${\cal O}$ with the same method i.e. through the correlation function $M_{i,j}$. Due to symmetry of the hopping term to go left or right on the left, this implies
\begin{equation}
\langle {\cal O}_{i,i-1}^{xy-yx}\rangle=\langle c^{\dagger}_i c_{i-1} - c^{\dagger}_{i-1} c_i\rangle=0.
\end{equation}
The bipartite current fluctuations in region $A$ then corresponds to the variance of the operator ${\cal O}$ such that
\begin{equation}
{\cal F}_{{\cal O}(A)} = \left\langle \left(\sum_{i\in A} {\cal O}_{i,i-1}^{xy-yx}\right)^2\right\rangle.
\end{equation}
Introducing the form of ${\cal O}$ in terms of fermions then this leads to evaluate
\begin{equation}
{\cal F}_{{\cal O}(A)} = -\frac{1}{4}\sum_{i,j\in A} \langle (c^{\dagger}_i c_{i-1} - c^{\dagger}_{i-1} c_i)(c^{\dagger}_j c_{j-1} - c^{\dagger}_{j-1} c_j) \rangle.
\end{equation}
Applying Wick theorem, then we derive the identity
\begin{equation}
{\cal F}_{{\cal O}(A)} = \frac{1}{2}{\cal F}_{S_z(A)} + \frac{1}{2\pi^2}\sum_{i\neq j\pm 1;i,j=1}^x \frac{\sigma_{i-j}-1}{(i-j)^2 -1}.
\end{equation}
The difference between the two evaluations of ${\cal F}_{S_z(A)}$ and ${\cal F}_{{\cal O}(A)}$ come from the indices i.e. the second term implies $i\neq j\pm 1$. It is in fact possible to evaluate this sum too and reach the result
\begin{equation}
{\cal F}_{{\cal O}(A)} = \frac{1}{2}{\cal F}_{S_z(A)} + \frac{1}{2\pi^2} \frac{x(\sigma_0 -1)}{-1} + 2\sum_{k=2}^x \frac{(x-k)(\sigma_k-1)}{k^2-1}.
\end{equation}
From definitions above $\sigma_0=0$. Introducing the digamma function $\psi_0(x)$ such that $\psi_0(x\rightarrow +\infty) = \ln x- \frac{1}{2x} -\frac{1}{2x^2}$ and the result for ${\cal F}_{S_z(A)}$ then we find
\begin{eqnarray}
{\cal F}_{{\cal O}(A)} &=& \frac{1}{2\pi^2}(\ln x + 1 +\gamma + \ln 2)  \nonumber \\
&+&\frac{1}{2\pi^2}\left(x-x+\psi_0\left(\frac{x+1}{2}\right) - \psi_0\left(\frac{1}{2}\right)\right) \nonumber \\
&=&  \frac{1}{2\pi^2}(\ln x + 1 +\gamma+ \ln2) \nonumber \\
&+& \frac{1}{2\pi^2}(\ln x -\ln 2+\gamma+2\ln 2).
\end{eqnarray}
It is particularly meaningful to see that this indeed simplifies and we reproduce the dominant logarithmic result from the Luttinger liquid theory i.e.
\begin{equation}
{\cal F}_{{\cal O}(A)} = \ln x+f_2 = \ln x + \frac{1}{2}+\gamma+\ln 2.
\end{equation}

\begin{center}
\begin{figure}[]
\includegraphics[width=16.5cm]{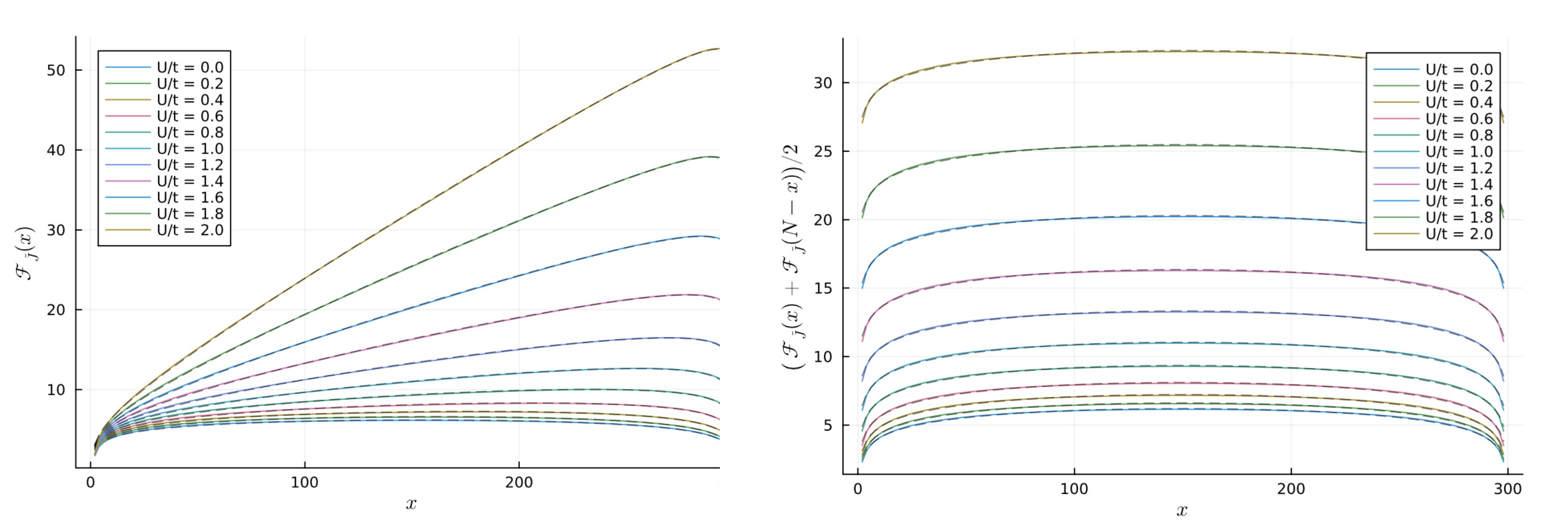}
\caption{(Left) Current fluctuations for a subsystem of size $x$ within a system of size $L=300$ with various values of $U/t$.  Solid colored lines are datas and fitting curves are the dashed lines. The fitting protocol is done with $f(x)=\frac{1}{\pi^2 K}\log d(x|L)+bx+cst$. The function $d(x|L)$ is the chord distance: 
$d(x|L)=\frac{L}{\pi}\sin\left(\frac{\pi x}{L}\right)$. (Right) Symmetrized bipartite current fluctuations when summing ${\cal F}_{\tilde{J}}(x)$ and ${\cal F}_{\tilde{J}}(N-x)$, with $N$ being the number of sites or length when fixing the lattice spacing to unity. We verify the fitting form in dashed grey lines,
$\frac{1}{\pi^2 K}\log L+cst.$ of the data (solid colored lines).}
\label{Figure}
\end{figure}
\end{center}

We verify these formulas with the DMRG approach and for the {\it interacting} case, i.e. with the XXZ model we find 
\begin{eqnarray}
{\cal F}_{S_z(A)} = \langle (\phi(x)-\phi(0))^2\rangle = K {\cal F}_1(x) \\ \nonumber
{\cal F}_{{\cal O}(A)} = \langle (\theta(x)-\theta(0))^2\rangle = K^{-1} {\cal F}_1(x).
\end{eqnarray}
The microscopic parameters $t$ and $U$ of the fermion model are related to the sound velocity $u$ and to the Luttinger parameter $K$ \cite{Haldane,Giamarchi}
\begin{eqnarray}
uK &=& v_F = 2t \\ \nonumber
\frac{u}{K} &=& v_F+\pi U = 2t+\pi U.
\end{eqnarray}
This is equivalent to 
\begin{equation}
K = \sqrt{\frac{2t}{2t+\pi U}} \hskip 1cm u = \sqrt{2t(2t+\pi U)}.
\end{equation}
The Bethe Ansatz form of the Luttinger parameter rather leads to \cite{Nishimoto}
\begin{equation}
K = \frac{\pi}{2 \hbox{arcos}\left(-\frac{U}{2t}\right)}.
\end{equation}
The two results for $K$ remain relatively close even if $U\sim 2t$ where $K\rightarrow \frac{1}{2}$. Within DMRG, first we verify the Luttinger parameter from the bipartite charge fluctuations \cite{chargefluctuations} with a very good agreement with Bethe Ansatz.
The current fluctuations are then evaluated showing a logarithmic plus a linear term. A proper protocol to subtract the linear term is shown in Fig. \ref{Figure}. We verify that the results remain identical for system sizes from $\sim 200$ to $\sim 400$, and they are clearer to read with Periodic Boundary Conditions (PBC). Then, we obtain the fractional charges shown in the Letter adding the logarithmic term for bipartite charge and current fluctuations $\pi^2({\cal F}_{Q(x)}+{\cal F}_{{\tilde{J}}(x)})$ which agree with Bethe Ansatz even when $U>t$. Increasing $U$, for $U=2t$ the Mott transition modifies both charge and current fluctuations, charge fluctuations become diminished and current fluctuations develop an enhanced linear term. 

\onecolumngrid

\twocolumngrid

\section{Bound State Solution of Jackiw-Rebbi Model, Quantum Field Theory and DMRG}
\label{AppendixB}

Here, we present our derivation of the bound state solution of the Jackiw-Rebbi model of interfaces \cite{JackiwRebbi,Shenbook} from a correspondence to a quantum field theory model describing a one-dimensional superconducting 
wire e.g. a Luther-Emery model \cite{KarynEPL,LutherEmery,FabrizioGogolin} with an additional zero-energy state.

We do a shift or translation of variables such that the interface is at $x=0$, the system keeps a length $L$. For $U=0$, the 1D wire is described through the kinetic term similar to a Dirac Hamiltonian \cite{Haldane,Giamarchi}
\begin{equation}
H_0 = -i v_F\int_{-\frac{L}{2}}^{+\frac{L}{2}} dx(c^{\dagger}_R \partial_x c_R - c^{\dagger}_L \partial_x c_L).
\end{equation}
The right-moving and left-moving fermions are introdudced in Eq. (\ref{electronoperator}). This Hamiltonian is obtained when linearizing the spectrum around half-filling. 
We include the smoothly varying potential term around $x=0$ which produces additional backscattering effects around the interface similar to the Jackiw-Rebbi model \cite{JackiwRebbi}
\begin{equation}
H_{potential} = \int_{-\frac{L}{2}}^\frac{L}{2} dx \left(\Delta \tanh\left(x\right) c^\dagger_L(x) c_R(x) +h.c.\right).
\end{equation}
Around $x=0$ this is the main effect of the barrier of potential. Similar to the preceding Section, the lattice spacing is set to unity for simplicity. 
The potential profile becomes flat away from the interface i.e. in this model we assume that the main source of potential is around the interface.
We can yet write the potential term under an integral from $-\frac{L}{2}$ to $+\frac{L}{2}$. To build a correspondence with the model in Refs. \cite{KarynEPL,FabrizioGogolin} we introduce
{\it one fermion operator} such that $c_R(x)=c(x)$ and $c_L(x)=-ic(-x)$. 

\begin{figure}[ht]
\includegraphics[width=10cm]{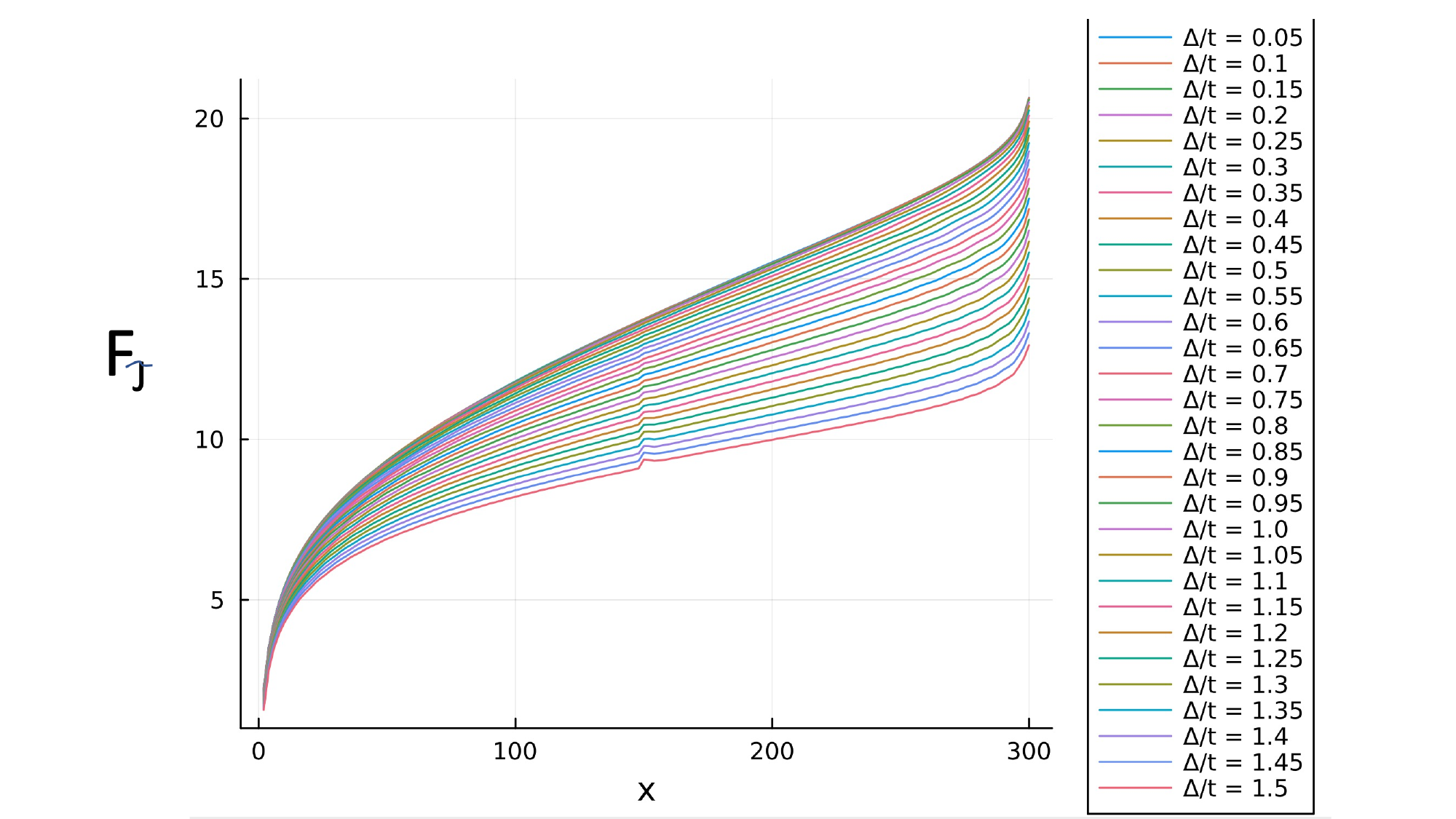}
\caption{Bipartite Current Fluctuations for Open Boundary Conditions (OBC) at $U=t$ with $300$ sites, 20 sweeps, for various values of $\Delta$.}
\label{currentfluctuations}
\end{figure}

\onecolumngrid
The potential term then precisely takes the form
\begin{equation}
H_{potential} = \int_{-\frac{L}{2}}^\frac{L}{2} dx \left(-i\Delta \tanh\left(x\right) c^{\dagger}(x)c(-x) + i\Delta \tanh\left(x\right) c^{\dagger}(-x)c(x)\right).
\end{equation}
Through the change of variable in the second term $x\rightarrow -x$
\begin{eqnarray}
\label{potential}
H_{potential} &=& -\int_{-\frac{L}{2}}^\frac{L}{2} dx \left(i\Delta \tanh\left(x\right) c^{\dagger}(x)c(-x) + i\Delta \tanh\left(x\right) c^{\dagger}(x)c(-x)\right). \\ \nonumber
&=& -\int_{-\frac{L}{2}}^\frac{L}{2} dx \left(2i\Delta  \tanh\left(x\right) c^{\dagger}(x) c(-x)\right).
\end{eqnarray}

\twocolumngrid
From bosonization of the operator $c^\dagger_L(x) c_R(x)$ this requires in principle to add Klein factors
$\eta_R \eta_L$ ensuring proper anti-commutation relations between the two fermions $c_L$ and $c_R$, which then results in an additional prefactor
$\pm i$ \cite{Haldane,Giamarchi}. This justifies why when introducing one fermion operator $c(x)$ along the whole path, $c^{\dagger}_R c_L$ turns into $i c^{\dagger}(x)c(-x)$. In principle, we can add any phase factor such that
\begin{equation}
c_L=e^{i\delta}c_R, 
\end{equation}
where $\delta=0$ corresponds to a perfect transmission i.e. no potential and $\delta=\pi$ to a complete backscattering with $c_R+c_L=0$ at $x=0$. The situation of a phase with $\delta=\frac{\pi}{2}$ indeed corresponds to an intermediate situation and the sign $\pm i$ in front of each term in Eq. (\ref{potential}) is fixed such that we have a normalized wave-function solution at $x=0$ instead of a diverging solution. Since the energy term due to backscattering effects is real then this requires to have an additional $i$ in front of the Hamiltonian density to compensate for this definition. We also have
\begin{equation}
H_0 =  -2i v_F\int_{-\frac{L}{2}}^{+\frac{L}{2}} dx \left(c^{\dagger}(x) \partial_x c(x)\right).
\end{equation}

To derive the zero-energy (electron-like) solution, we introduce the energy-resolved creation operator
\begin{equation}
c_{\epsilon}^{\dagger} = \int_{-\frac{L}{2}}^{\frac{L}{2}} dx \chi_{\epsilon}(x) c^{\dagger}(x),
\end{equation}
such that
\begin{equation}
H = H_0 +H_{potential} = \sum_{\epsilon} \epsilon c^{\dagger}_{\epsilon} c_{\epsilon}.
\end{equation}
Playing with commutation relations $[c_{\epsilon},c^{\dagger}_{\epsilon'}] = \{c_{\epsilon},c^{\dagger}_{\epsilon'}\} - 2 c^{\dagger}_{\epsilon'} c_{\epsilon} = \delta_{\epsilon \epsilon'}-2 c^{\dagger}_{\epsilon'} c_{\epsilon}$ we verify the equation
\begin{equation}
[H,c^{\dagger}_{\epsilon}] = \epsilon c^{\dagger}_{\epsilon} = \int_{-\frac{L}{2}}^{\frac{L}{2}} dx \epsilon \chi_{\epsilon}(x) c^{\dagger}(x)
\end{equation}
which is indeed the solution of an eigenstate with energy $\epsilon$. From the form of $H=H_0+H{potential}$ we also have

\onecolumngrid
\begin{equation}
[H,c^{\dagger}_{\epsilon}] = -2i v_F\int_{-\frac{L}{2}}^{\frac{L}{2}} dx c^{\dagger}(x)\partial_x c(x) - 2i\Delta\int_{-\frac{L}{2}}^{\frac{L}{2}} dx \tanh(x) c^{\dagger}(x) \chi_{\epsilon}(-x).
\end{equation}
\twocolumngrid
This is equivalent to
\begin{equation}
\epsilon \chi_{\epsilon}(x) = -2 i v_F \partial_x \chi_{\epsilon}(x) - 2i\Delta \tanh(x) \chi_{\epsilon}(-x).
\end{equation}
To find the solution at zero energy we set $\epsilon=0$. A symmetric bound state around the interface satisfies the general solution $\chi_0(x)={\cal A} e^{-f(|x|)}$. We can verify that the equation is satisfied for $x>0$ and equally for $x<0$. Suppose $x>0$
then the solution leads to $f'(x)=\frac{\Delta}{v_F}\tanh x$ i.e. $f(x)=\frac{\Delta}{v_F}\ln(\cosh(x))$. Around $x=0$, the function $f(x)$ can be simplified to $f(x)\sim \frac{\Delta}{2v_F} x^2$ with $v_F=2t$ such that normalization of the wavefunction for 
$L\rightarrow +\infty$ leads
to 
\begin{equation}
\chi_0(x) = \left(\frac{\Delta}{2t\pi}\right)^{\frac{1}{4}} e^{-\frac{\Delta}{2t}\ln\cosh|x|}.
\end{equation}
We emphasize here that in the presence of a jump of potential at an edge, the bound state remains e.g. with a $sgn(x)$ function in front of the potential term but the form of the wave-function $\chi_0(x)$ is slightly modified \cite{FabrizioGogolin}.
The $H_{potential}$ term in the Luther-Emery model corresponds to a backscattering term in the spin sector coming from an s-wave pairing interaction \cite{LutherEmery,FabrizioGogolin}. 
The model is generalizable to a d-wave superconductor in a ladder \cite{KarynEPL}.  Coupling this zero-energy state to a magnetic impurity can then give rise to zero-energy Majorana fermions \cite{KarynEPL,Karyn2}.

\begin{figure}[t]
\includegraphics[width=10cm]{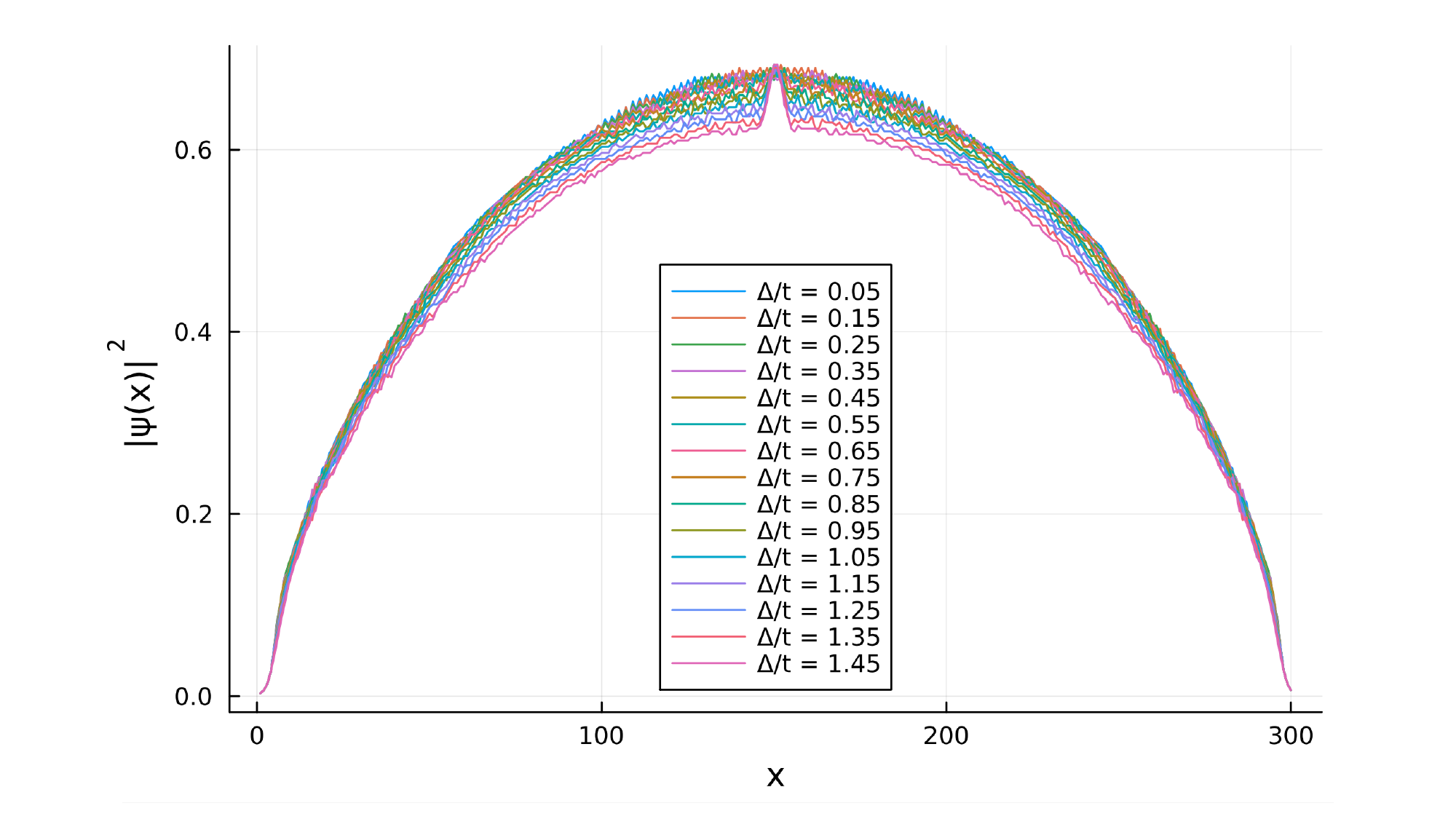}
\caption{Probability density (not normalized) for OBC at $U=t$ with $300$ sites, 20 sweeps, for various values of $\Delta$.}
\label{probabilityUt1}
\end{figure}

In Figs. \ref{currentfluctuations} and \ref{probabilityUt1}, we present additional material on DMRG results showing how the signals for the density probability and the bipartite current fluctuations remain visible for intermediate values of interactions $U\sim t$. We verify the occurrence of the bound state in the density probability until $U\sim 1.75t$. 

\end{appendix}


\begin{thebibliography}{9}

\bibitem{Haldane}
F. D. M. Haldane, 'Luttinger liquid theory' of one-dimensional quantum fluids. I. Properties of the Luttinger model and their extension to the general 1D interacting spinless Fermi gas, J. Phys. C: Solid State Phys. {\bf 14} 2585 (1981).

\bibitem{Giamarchi}
T. Giamarchi, Quantum Physics in One Dimension, Clarendon Press, Oxford 2003.

\bibitem{PhamOrsay}
K. V. Pham, M. Gabay and P. Lederer, Fractional excitations in the Luttinger liquid, Phys. Rev. B {\bf 61}, 16397 (2000).

\bibitem{Laughlin}
R. B. Laughlin, Anomalous Quantum Hall Effect: An Incompressible Quantum Fluid with Fractionally Charged Excitations, Phys. Rev. Lett. {\bf 50}, 1395 (1983).

\bibitem{SafiSchulz}
I. Safi and H. J. Schulz, "Correlated Fermions and Transport in Mesoscopic Systems", edited by T. Martin, G. Montambaux and J. Tran Thanh Van. Editions Frontieres, Gif-sur-Yvette (1996); arXiv:cond-mat/9711082.

\bibitem{MaslovStone}
D. L. Maslov and M. Stone, Landauer Conductance of Luttinger Liquids with Leads, Phys. Rev. B {\bf 52}, R5539(R) (1995).

\bibitem{Tarucha}
S. Tarucha, T. Honda, and T. Saku, Reduction of quantized conductance at low temperatures observed in $2$ to $10 \mu m$-long quantum wires, Solid State Commun. {\bf 94}, 413 (1995).

\bibitem{Steinberg}
H. Steinberg, G. Barak, A. Yacoby, L. N. Pfeiffer, K. W. West, B. I. Halperin and Karyn Le Hur, Nature Physics {\bf 4}, 116-119 (2008).

\bibitem{KarynBertAmir}
K. Le Hur, B. I. Halperin and A. Yacoby, Annals of Physics {\bf 323}, 3037-3058 (2008).

 \bibitem{Karyn}
 K. Le Hur, Electron fractionalization induced dephasing in Luttinger liquids, Phys. Rev. B {\bf 65}, 233314 (2002); K. Le Hur, Dephasing from electron fractionalization, Phys. Rev. Lett. {\bf 95}, 076801 (2005); K. Le Hur, The electron lifetime in Luttinger liquids,  Phys. Rev. B {\bf 74}, 165104 (2006).
 
 \bibitem{Rice}
 W. Apel and T. M. Rice, Localisation and interaction in one dimension, J. Phys. C: Solid State Phys. {\bf 16} L271 (1983).
 
 \bibitem{Hansen}
 A. E. Hansen, A. Kristensen, S. Pedersen, C. B. Sorensen, and P. E. Lindelof Phys. Rev. B {\bf 64}, 045327 (2001).
 
 \bibitem{SeeligButtiker}
G. Seelig and M. B\"uttiker, Charge-fluctuation-induced dephasing in a gated mesoscopic interferometer, Phys. Rev. B {\bf 64}, 245313 (2001).

\bibitem{Kapfer}
M. Kapfer, P. Roulleau, M. Santin, I. Farrer, D. A. Ritchie and D. C. Glattli, A Josephson relation for fractionally charged anyons, Science {\bf 363}, Issue 6429
pp. 846-849 (2019).

\bibitem{Glattli}
L. Saminadayar, D. C. Glattli, Y. Jin, and B. Etienne, Observation of the e/3 Fractionally Charged Laughlin Quasiparticles, Phys. Rev. Lett. {\bf 79}, 2526 (1997).

\bibitem{Picciotto}
 R. de-Picciotto, M. Reznikov, M. Heiblum, V. Umansky, G. Bunin and D. Mahalu, Direct observation of a fractional charge, Nature {\bf 389}, 162-164 (1997).
 
 \bibitem{TalKaryn}
 T. Goren and K. Le Hur, Real-Time Ramsey Interferometry in Fractional Quantum Hall States, Phys. Rev. B {\bf 99}, 161109 (2019).
 
\bibitem{Lin}
 Z.-K. Lin, Y. Zhou, B. Jiang, B.-Q. Wu, L.-M. Chen, X.-Y. Liu, L.-W. Wang, P. Ye and J.-H. Jiang, Measuring entanglement entropy and its topological signature for phononic systems, Nature Communications, {\bf 15} 1601 (2024).

\bibitem{Harvard}
R. Islam, R. Ma, Ph. M. Preiss, M. E. Tai, A. Lukin, M. Rispoli and M. Greiner, Measuring entanglement entropy through the interference of quantum many-body twins, Nature {\bf 528}, 77-83 (2015).
A. Lukin, M. Rispoli, R. Schittko, M. E. Tai, A. M. Kaufman, S. Choi, V. Khemani, J. L\' eonard and M. Greiner, Probing entanglement in a many-body-localized system, Science {\bf 364}, Issue 6437, 256-260 (2019).

\bibitem{CalabreseCardy}
P. Calabrese and J. Cardy, Entanglement entropy and conformal field theory, J.Phys. A {\bf 42} 504005 (2009).

\bibitem{chargefluctuations}
H. F. Song, S. Rachel, and K. Le Hur, General relation between entanglement and fluctuations in one dimension, Phys. Rev. B {\bf 82}, 012405 (2010); {\it Review:} H. F. Song, S. Rachel, C. Flindt, I. Klich, N. Laflorencie, and K. Le Hur, Bipartite fluctuations as a probe of many-body entanglement, Phys. Rev. B {\bf 85}, 035409 (2012) Editors' Suggestion;  I. Klich and L. Levitov, Quantum Noise as an Entanglement Meter, Phys. Rev. Lett. {\bf 102}, 100502 (2009); H. Oshima and  Y. Fuji, Charge fluctuation and charge-resolved entanglement in a monitored quantum circuit with $U(1)$-symmetry Phys. Rev. B {\bf 107}, 014308 (2023).

\bibitem{Fradkin}
B. Hsu, E. Grosfeld, and E. Fradkin, Quantum noise and entanglement generated by a local quantum quench, Phys. Rev. B {\bf 80}, 235412 (2009).

\bibitem{FQHE1/2}
 A. Petrescu, M. Piraud, G. Roux, I. P. McCulloch and K. Le Hur, Precursor of Laughlin state of hard core bosons on a two leg ladder, Phys. Rev. B {\bf 96}, 014524 (2017).
 
 \bibitem{Berg}
E. Berg, Y. Oreg, E.-A. Kim and F. von Oppen, Fractional Charges on an Integer Quantum Hall Edge, Phys. Rev. Lett. {\bf 102}, 236402 (2009).

\bibitem{Gutman}
D. B. Gutman, Y. Gefen and A. D. Mirlin, Full Counting Statistic of a Luttinger Liquid Conductor, Phys. Rev. Lett. {\bf 105}, 256802 (2010).

\bibitem{Biswas}
S. Biswas, R. Bhattacharyya, H. K. Kundu, A. Das, M. Heiblum, V. Umansky, M. Goldstein and Y. Gefen, Nature Physics {\bf 18}, 1476 (2022).

\bibitem{Herviou}
Lo\" ic Herviou, C. Mora, and K. Le Hur, Bipartite charge fluctuations in one-dimensional Z2 superconductors and insulators, Phys. Rev. B {\bf 96}, 121113(R) (2017).

\bibitem{Konik}
R. K. Malla, A. Weichselbaum, T.-C. Wei and R. M. Konik, Detecting Multipartite Entanglement Patterns using Single Particle Green’s Functions, Phys. Rev. Lett. {\bf 133}, 260202 (2024).

\bibitem{Tennant}
P. Laurell, A. Scheie, C. J. Mukherjee, M. M. Koza, M. Enderle, Z. Tylczynski, S. Okamoto, R. Coldea, D. Alan Tennant, and G. Alvarez, Quantifying and Controlling Entanglement in the Quantum Magnet Cs2CoCl4,
Phys. Rev. Lett. {\bf 127}, 037201 (2021).

\bibitem{Affleck}
I. Affleck, Les Houches Summer School in Theoretical Physics 1988: Fields, Strings, Critical Phenomena, editors E. Brezin and J. Zinn-Justin, North-Holland, 1990.

\bibitem{chargecurrent}
A. Chatterjee, S. D. Pace, S.-H. Shao, Quantized axial charge of staggered fermions and the chiral anomaly, Phys. Rev. Lett. {\bf 134}, 021601 (2025).

\bibitem{Nishimoto}
S. Nishimoto, Tomonaga-Luttinger-liquid criticality: numerical entanglement entropy approach, Phys. Rev. B {\bf 84}, 195108 (2011).

\bibitem{ITensors}
The Density Matrix Renormalization Group results are obtained from ITensors and ITensors MPS: https://itensor.org/.

\bibitem{datajustification}
We add a link such that our data for the Figures are accessible: https://zenodo.org/records/14899480. 
A link to the developed algorithms may also be provided if necessary; we emphasize that the DMRG codes are adapted from https://itensor.org/. 

\bibitem{fluctuationsT}
S. Rachel, N. Laflorencie, H. F. Song and K. Le Hur, Detecting Quantum Critical Points Using Bipartite Fluctuations, Phys. Rev. Lett. {\bf 108}, 116401 (2012).

\bibitem{JackiwRebbi}
R. Jackiw and C. Rebbi, Solitons with fermion number, Phys, Rev. D {\bf 13}, 3398 (1976).

\bibitem{Shenbook}
Shun-Qing Shen: Chap 2. of Topological insulators: Dirac equation in condensed matter. 2nd
edition. (Springer series in solid-state science 187, Berlin, 2017). Chapter 2. (2017).

\bibitem{Angelakis}
D. G. Angelakis, P. Das and C. Noh, Probing the topological properties of the Jackiw-Rebbi model with light, Scientific Reports {\bf 4}, Article number: 6110 (2014). 

\bibitem{KarynEPL}
K. Le Hur, Kondo effect in a one-dimensional d-wave superconductor, EPL {\bf 49} 768 (2000).

\bibitem{LutherEmery}
A. Luther and V. J. Emery, Backward Scattering in the One-Dimensional Electron Gas, Phys. Rev. Lett. {\bf 33}, 589 (1974).

\bibitem{FabrizioGogolin}
 M. Fabrizio and A. O. Gogolin, Interacting one dimensional electron gas with open boundaries, Phys. Rev. B {\bf 51}, 17827 (1995).

\bibitem{pwavearticle}
L. Herviou, C. Mora and K. Le Hur, Phase diagram and entanglement of two interacting topological Kitaev chains, Phys. Rev. B {\bf 93}, 165142 (2016).

\bibitem{Kitaev}
A. Kitaev, Unpaired majorana fermions in quantum
wires, Physics Uspekhi {\bf 44}, 131 (2001).

\bibitem{correlations}
C. S. Chiu, G. Ji, A. Bohrdt, M. Xu, M. Knap, E. Demler, F. Grusdt, M. Greiner and D. Greif, String patterns in the doped Hubbard model, Science {\bf 365} 251-256 (2019).
 
 \bibitem{chargefluctuationsgate}
 D. Berman, N. B. Zhitenev, R. C. Ashoori and M. Shayegan, Observation of Quantum Fluctuations of Charge on a Quantum Dot, Phys. Rev. Lett. {\bf 82}, 161 (1999).
 
 \bibitem{Karyn2}
 K. Le Hur, Free Majorana Modes in Superconducting Quantum Wires, arXiv:2511.03380.
 
 
\end{thebibliography}
\end{document}